\documentclass[journal]{IEEEtran}

\usepackage{latexsym}
\usepackage{graphics}
\usepackage{epsfig,amssymb,amsmath,graphicx,amsfonts}
\usepackage[dvips]{color}
\usepackage{cite,hyperref}

\newtheorem{theorem}{Theorem}

\newtheorem{lemma}{Lemma}
\newtheorem{proposition}{Proposition}

\begin{document}

\title{Secret Writing on Dirty Paper: A Deterministic View}

\author{Mustafa El-Halabi, Tie Liu, Costas Georghiades, and Shlomo Shamai (Shitz)
\thanks{This research was supported in part by the National Science Foundation under Grant
CCF-08-45848, by the European Commission in the framework of the FP7 Network of Excellence in
Wireless Communications NEWCOM++, and by the Israel Science Foundation.
}
\thanks{Mustafa El-Halabi, Tie Liu, and Costas Georghiades are with the Department of Electrical and Computer Engineering,
Texas A\&M University, College Station, TX 77843, USA (email: \{mustafa79,tieliu,c-georghiades\}@tamu.edu)}
\thanks{Shlomo Shamai (Shitz) is with the Department of Electrical Engineering, Technion -- Israel Institute of Technology, Technion City, Haifa 32000, Israel (email: sshlomo@ee.technion.ac.il)}
}

\maketitle

\begin{abstract}
Recently there has been a lot of success in using the deterministic approach to provide approximate characterization of Gaussian network capacity. In this paper, we take a deterministic view and revisit the problem of wiretap channel with side information. A precise characterization of the secrecy capacity is obtained for a linear deterministic model, which naturally suggests a coding scheme which we show to achieve the secrecy capacity of the degraded Gaussian model (dubbed as ``secret writing on dirty paper") to within half a bit.
\end{abstract}

\begin{IEEEkeywords}
Dirty-paper coding, information-theoretic security, linear deterministic model, side information, wiretap channel  
\end{IEEEkeywords}

\section{Introduction}
In information theory, an interesting and useful communication model is a state-dependent channel where the channel states are non-causally known at the transmitter as side information. Of particular importance is a discrete-time channel with real input and additive white Gaussian noise and interference, where the interference is non-causally known at the transmitter as side information. 

Costa \cite{Cos-IT83} was the first to study this communication scenario, which he whimsically coined as ``writing on dirty paper." Based on an earlier result of Gel'fand and Pinsker \cite{GP-PCIT80}, Costa \cite{Cos-IT83} proved the surprising result that the capacity of writing on dirty paper is the \emph{same} as that of writing on clean paper without interference. Since \cite{Cos-IT83}, dirty-paper coding has found a wide range of applications in digital watermarking and network communications, particularly involving broadcast scenarios.

Recent works \cite{MVL-IT06} and \cite{CV-IT08} studied the problem of dirty-paper coding in the presence of an additional eavesdropper, which is a natural extension of Costa's dirty-paper channel to the secrecy communication setting. In this scenario, which we dub as ``secret writing on dirty paper", the legitimate receiver channel is a dirty-paper channel of Costa. The signal received at the eavesdropper, on the other hand, is assumed to be a \emph{degraded} version of the signal received at the legitimate receiver.  An achievable secrecy rate was established based on a double-binning scheme and was shown to be the secrecy capacity of the channel under some channel parameter configurations \cite{MVL-IT06,CV-IT08}. For the \emph{general} channel parameter configuration, however, the secrecy capacity of the channel remains \emph{unknown}. 

In facing some challenging Gaussian network communication problems, recent advances \cite{ETW-IT08,ADT-IT10} in network information theory advocate a \emph{deterministic} approach and seeks \emph{approximate} characterization of the network capacity to within \emph{finite} bits (regardless of the received signal-to-noise ratios). Motivated by the success of \cite{ETW-IT08} and \cite{ADT-IT10}, in this paper we take a deterministic view and revisit the problem of wiretap channel with side information. A precise characterization of the secrecy capacity is obtained for a linear deterministic model, which naturally suggests a coding scheme which we show to achieve the secrecy capacity of the degraded Gaussian model to within half a bit.

The rest of the paper is organized as follows. In Sec.~\ref{sec:dpc}, we first take a deterministic view at Costa's dirty-paper channel and provide an approximate characterization of the channel capacity to within half a bit. Note that even though a precise characterization of Costa's dirty-paper channel is well known \cite{Cos-IT83}, the proposed approximate characterization establishes a framework for studying side-information problems via the deterministic approach. Building on the framework of Sec.~\ref{sec:dpc}, in Sec.~\ref{sec:sdpc} we extend the deterministic approach to the problem of secret writing on dirty paper and provide an approximate characterization of the secrecy capacity to within half a bit. A different but closely related communication scenario known as secret-key agreement via dirty-paper coding is discussed in Sec.~\ref{sec:kdpc}. Finally, in Sec.~\ref{sec:con} we conclude the paper with some remarks.
 
\section{Writing on Dirty Paper}\label{sec:dpc}
\subsection{Gaussian Model}
Consider the dirty-paper channel of Costa \cite{Cos-IT83}, where the received signal $Y[i]$ at time index $i$ is given by
\begin{equation}
Y[i]=hX[i]+gS[i]+N[i].
\label{eq:ch-dpc}
\end{equation}
Here, $X[i]$ is the channel input which is subject to a \emph{unit} average power constraint, $N[i]$ and $S[i]$ are independent \emph{standard} Gaussian noise and interference and are independently identically distributed (i.i.d.) across the time index $i$, and $h$ and $g$ are the (real) channel coefficients corresponding to the channel input and interference, respectively. The interference $S[i]$ is assumed to be non-causally known at the transmitter as side information. The channel coefficients $h$ and $g$ are fixed during communication and are assumed to be known at both transmitter and receiver. 

The channel capacity, as shown by Costa \cite{Cos-IT83}, is given by
\begin{equation*}
C=I(U;Y)-I(U;S)
\end{equation*}
where the input variable $X$ is standard Gaussian and independent of the known interference $S$, and $U$ is an auxiliary variable chosen as
\begin{equation}
U=hX+\frac{h^2}{h^2+1}gS.
\label{eq:aux}
\end{equation}
For this choice of auxiliary-input variable pair $(U,X)$, 
\begin{equation*}
I(U;Y)-I(U;S)=\frac{1}{2}\log(1+h^2)
\end{equation*}
which equals the capacity of the channel \eqref{eq:ch-dpc} when the interference $S[i]$ is also known at the receiver.

\subsection{Linear Deterministic Model}
Consider the linear deterministic model \cite{ADT-IT10} for Costa's dirty-paper channel \eqref{eq:ch-dpc}, where the received signal $Y[i]$ at time index $i$ is given by
\begin{equation}
Y[i]=D^{q-n}X[i] \oplus D^{q-m}S[i].
\label{eq:ch-ldpc}
\end{equation}
Here, $X[i]$ is the binary input vector of length $q=\max\{n,m\}$, $S[i]$ is the i.i.d. interference vector whose elements are i.i.d. Bernoulli-$1/2$, $D=[d_{j,k}]$ is the $q\times q$ down-shift matrix with elements
\begin{equation*}
d_{j,k} = \left\{ \begin{array}{ll} 1 & \mbox{if} \; 2 \leq j=k+1 \leq q\\
0 & \mbox{otherwise}
\end{array} \right.
\end{equation*}
and $n$ and $m$ are the integer channel gains corresponding to the channel input and interference, respectively. The vector interference $S[i]$ is assumed to be non-causally known at the transmitter as side information. The channel gains $n$ and $m$ are fixed during communication and are assumed to be known at both transmitter and receiver. 

Following the result of Gel'fand and Pinsker \cite{GP-PCIT80}, the capacity of the linear deterministic dirty-paper channel  \eqref{eq:ch-ldpc} is given by
\begin{equation*}
C=I(U;Y)-I(U;S)
\end{equation*}
where the input variable $X$ is an i.i.d. Bernoulli-$1/2$ random vector and independent of $S$, and $U$ is an auxiliary variable chosen as
\begin{equation}
U=Y=D^{q-n}X\oplus D^{q-m}S.
\label{eq:laux}
\end{equation}
For this choice of the auxiliary-input variable pair $(U,X)$,
\begin{eqnarray*}
I(U;Y)-I(U;S) & = & I(Y;Y)-I(Y;S)\\
&=& H(Y)-I(Y;S)\\
& = & H(Y|S)\\
&=& H(D^{q-n}X)\\
&=& rank(D^{q-n})\\
&=& n
\end{eqnarray*}
which equals the capacity of the channel \eqref{eq:ch-ldpc} when the interference $S[i]$ is also known at the receiver.

We emphasize that in \eqref{eq:laux}, we may choose $U=Y$ only because $Y$ here is a \emph{deterministic} function of $X$ and $S$. In fact, for \emph{any} deterministic Gel'fand-Pinsker channel (not necessarily linear) where the channel output $Y$ is a deterministic (bivariate) function of the channel input $X$ and state $S$, 
$\max_{p(x|s)}H(Y|S)$
is the capacity of the channel when the channel state $S$ is also known at the receiver. Thus, $U=Y$ is always an optimal choice for deterministic Gel'fand-Pinsker channels, a fact which was also observed in \cite{KM-ISIT11} recently. 

\subsection{Connections between the Gaussian and the Linear Deterministic Model}
A quick comparison between the Gaussian \eqref{eq:ch-dpc} and the linear deterministic \eqref{eq:ch-ldpc} models reveals the following equivalence relationship between these two models:
\begin{equation}
h \longleftrightarrow D^{q-n} \quad \mbox{and} \quad g \longleftrightarrow D^{q-m}.
\label{eq:eqv}
\end{equation}
Given this equivalence relationship, the optimal choice \eqref{eq:laux} of auxiliary variable $U$ for the linear deterministic model \eqref{eq:ch-ldpc} naturally suggests the following choice of auxiliary variable $U$ for the Gaussian model \eqref{eq:ch-dpc}:
\begin{equation}
U=hX+gS
\label{eq:aux2}
\end{equation} 
where $X$ is standard Gaussian and independent of $S$. Compared with the optimal choice \eqref{eq:aux}, the choice \eqref{eq:aux2} of auxiliary variable $U$ is \emph{suboptimal}. However, for this suboptimal choice of auxiliary-input variable pair $(U,X)$,
\begin{eqnarray*}
I(U;S) &=& \frac{1}{2}\log\left(1+\frac{g^2}{h^2}\right)\\
\mbox{and} \quad I(U;Y) &=& \frac{1}{2}\log(1+h^2+g^2)
\end{eqnarray*}
giving an achievable rate
\begin{eqnarray*}
R &=& \left[I(U;Y)-I(U;S)\right]^+\\
&=& \left[\frac{1}{2}\log\frac{(1+h^2+g^2)h^2}{h^2+g^2}\right]^+\\
& \geq & \left[\frac{1}{2}\log(h^2)\right]^+
\end{eqnarray*}
which is always within half a bit of the actual channel capacity
$C=\frac{1}{2}\log(1+h^2)$. Here, we denote $x^+:=\max\{0,x\}$ so that the achievable rates are always nonnegative.

The fact that the choice \eqref{eq:aux2} of auxiliary variable $U$ leads to an achievable rate which is always within half a bit of the dirty-paper channel capacity is well known (see \cite{ZSE-IT02} for example). However, it is interesting to see that such a choice comes up \emph{naturally} in the context of the deterministic approach.

\section{Secret Writing on Dirty Paper}\label{sec:sdpc}
Having understood how the linear deterministic model of \cite{ADT-IT10} may be used to obtain an approximate characterization of the capacity of Costa's dirty-paper channel, next we shall extend the deterministic approach to the problem of secret writing on dirty paper. 

\subsection{Discrete Memoryless Model}
\begin{figure}[t!]
\centering
\includegraphics[width=\linewidth,draft=false]{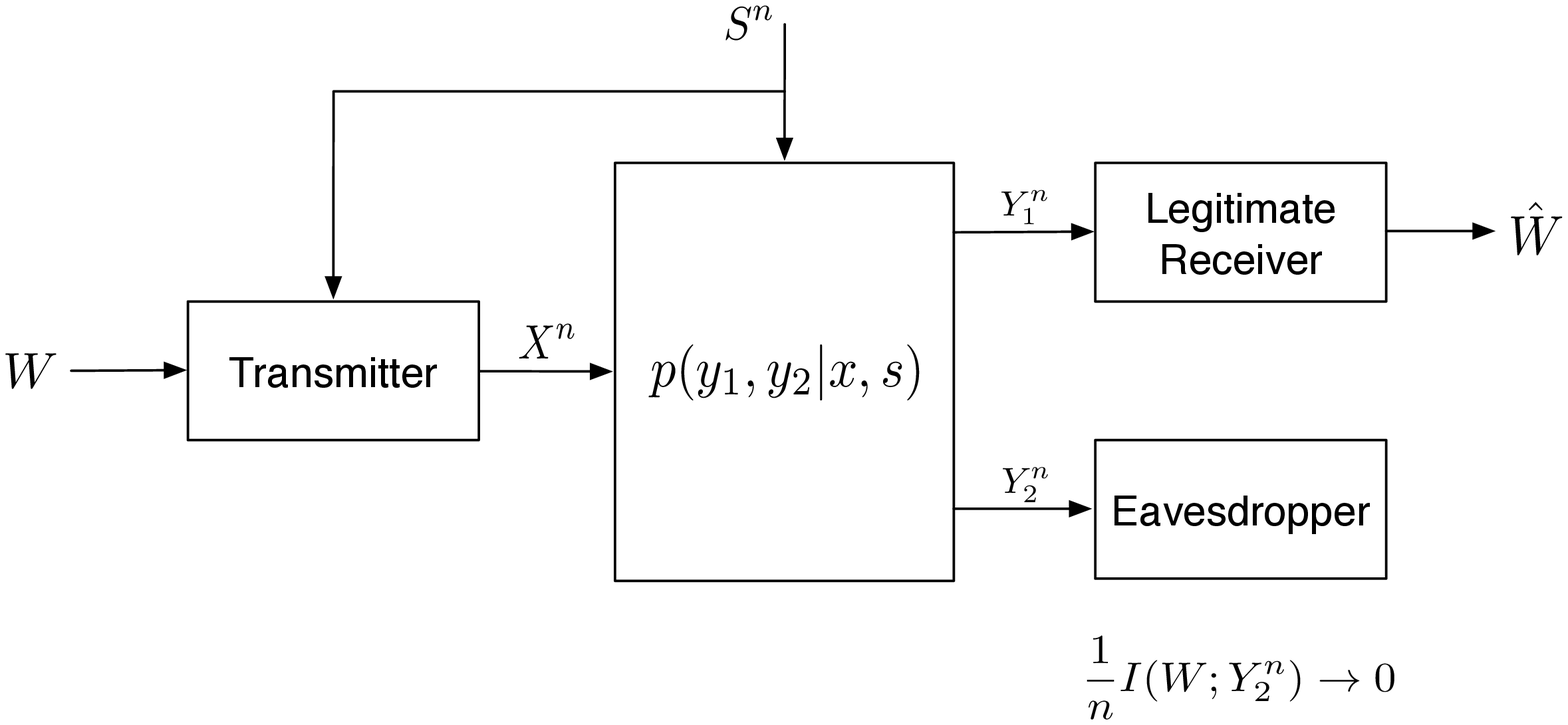}
\caption{Wiretap channel with side information.}
\label{fig}
\end{figure}

As illustrated in Fig.~\ref{fig}, consider a discrete-time memoryless wiretap channel with transition probability $p(y_1,y_2|x,s)$, where $X[i]$ is the channel input (at time index $i$), $S[i]$ is the channel state, and $Y_1[i]$ and $Y_2[i]$ are the received signals at the legitimate receiver and the eavesdropper, respectively. The channel state $S[i]$ is i.i.d. across the time index $i$ and is assumed to be non-causally known at the transmitter as side information. The transmitter has a message $W$, which is intended for the legitimate receiver but needs to be kept asymptotically perfectly secret from the eavesdropper. Following the classical works \cite{Wyn-BSTJ75} and \cite{CK-IT78}, it is required that
\begin{equation}
\frac{1}{n}I(W;Y_2^n) \rightarrow 0
\end{equation}
in the limit as the block length $n \rightarrow \infty$, where $Y_2^n:=(Y_2[1],\ldots,Y_2[n])$. The secrecy capacity $C_s$ is defined as the \emph{largest} secrecy rate that can be achieved by a coding scheme.  

Chen and Vinck \cite{CV-IT08} derived a single-letter lower bound on the secrecy capacity (an achievable secrecy rate), which can be written as
\begin{equation}
\begin{array}{l}
C_s \geq \max_{p(u,x|s)}\min\left\{I(U;Y_1)-I(U;S),\right.\\
\hspace{120pt} \left.I(U;Y_1)-I(U;Y_2)\right\}
\end{array}
\label{eq:lb}
\end{equation}
where $U$ is an auxiliary variable satisfying the Markov chain $U \rightarrow (X,S) \rightarrow (Y_1,Y_2)$.

We also have the following simple upper bound on the secrecy capacity. 

\begin{proposition} \label{lemma:ub}
The secrecy capacity $C_s$ of a discrete memoryless wiretap channel $p(y_1,y_2|x,s)$ with channel state $S$ non-causally known at the transmitter as side information can be bounded from above as
\begin{equation}
C_s \leq \max_{p(x|s)}\min\left\{I(X;Y_1|S),I(X,S;Y_1|Y_2)\right\}.
\label{eq:ub}
\end{equation}
\end{proposition}

Note that $\max_{p(x|s)}I(X;Y_1|S)$ is an upper bound on the Shannon capacity of the legitimate receiver channel by giving the channel state $S$ to the legitimate receiver, and $\max_{p(x|s)}I(X,S;Y_1|Y_2)$ is an upper bound on the secrecy capacity of the wiretap channel by allowing the transmit message $W$ to be encoded by the channel state $S$ (i.e., fully action-dependent state \cite{Wei-IT10}) and by giving the received signal $Y_2$ to the legitimate receiver. Here, a simple single-letterization technique of Willems \cite{Wil-ITB00} allows the maximizations to be moved outside the minimization. See Appendix~\ref{app1} for the details of the proof.

For \emph{semi-deterministic} channels where the channel output at the legitimate receiver is a deterministic (bivariate) function of the channel input and state, the lower \eqref{eq:lb} and the upper \eqref{eq:ub} bounds coincide, leading to a \emph{precise} characterization of the secrecy capacity. The result is summarized in the following theorem.

\begin{theorem}\label{thm:det}
Consider a discrete memoryless wiretap channel $p(y_1,y_2|x,s)$ with channel state $S$ non-causally known at the transmitter as side information. If the received signal $Y_1$ at the legitimate receiver is a \emph{deterministic} function of the channel input $X$ and state $S$, i.e., $Y_1=f(X,S)$ for some bivariate function $f$, the secrecy capacity $C_s$ of the channel is given by
\begin{equation}
C_s = 
\max_{p(x|s)}\min\left\{H(Y_1|S),H(Y_1|Y_2)\right\}.
\label{eq:bd}
\end{equation}
\end{theorem}

\begin{IEEEproof}
The fact that $$C_s \geq 
\max_{p(x|s)}\min\left\{H(Y_1|S),H(Y_1|Y_2)\right\}$$ follows from the lower bound \eqref{eq:lb} by setting $U=Y_1$ (we may do so only because here $Y_1$ is a deterministic function of $X$ and $S$), which gives 
\begin{equation*}
I(U;Y_1)-I(U;S) = H(Y_1)-I(Y_1;S)  = H(Y_1|S)
\end{equation*}
and similarly
$$I(U;Y_1)-I(U;Y_2)= H(Y_1)-H(Y_1|Y_2) = H(Y_1|Y_2).$$

The converse part of the theorem follows from the upper bound \eqref{eq:ub} and the fact that $Y_1$ is a deterministic function of $(X,S)$, so we have
\begin{equation*}
I(X;Y_1|S)=H(Y_1|S)-H(Y_1|X,S)=H(Y_1|S)
\end{equation*}
and
\begin{equation*}
I(X,S;Y_1|Y_2)=H(Y_1|Y_2)-H(Y_1|X,S,Y_2)=H(Y_1|Y_2).
\end{equation*}
This completes the proof of the theorem.
\end{IEEEproof}

Note that when the channel state $S$ is deterministic, a semi-deterministic wiretap channel with side information reduces to a regular semi-deterministic wiretap channel without side information. In this case, let $S$ be a constant in \eqref{eq:bd} and we have
\begin{equation*}
C_s = 
\max_{p(x)}\min\left\{H(Y_1),H(Y_1|Y_2)\right\} = \max_{p(x)}H(Y_1|Y_2)
\end{equation*}
which recovered the result of \cite{GVLP-ITW07} on the secrecy capacity of the semi-deterministic wiretap channel (without side information).

\subsection{Linear Deterministic Model}
Next, let us use the result of Theorem~\ref{thm:det} to determine the secrecy capacity of a linear deterministic wiretap channel with side information. In this model, the received signals (at time index $i$) at the legitimate receiver and the eavesdropper are given by
\begin{equation}
\begin{array}{lll}
Y_1[i] &=& D^{q-n_1}X[i] \oplus D^{q-m_1}S[i]\\
Y_2[i] &=& D^{q-n_2}X[i] \oplus D^{q-m_2}S[i]
\end{array}
\label{eq:ch-lsdpc}
\end{equation}
where $X[i]$ is the binary input vector of length $q=\max\{n_1,n_2,m_1,m_2\}$, $S[i]$ is the i.i.d. vector interference whose elements are i.i.d. Bernoulli-$1/2$, $D$ is the $q\times q$ down-shift matrix, and $n_1$, $n_2$, $m_1$ and $m_2$ are the integer channel gains. The vector interference $S[i]$ is assumed to be non-causally known at the transmitter as side information. The channel gains $n_1$, $n_2$, $m_1$ and $m_2$ are fixed during communication and are assumed to be known at all terminals. 

The following theorem provides an \emph{explicit} characterization of the secrecy capacity of the linear deterministic wiretap channel \eqref{eq:ch-lsdpc} with side information.

\begin{theorem}\label{thm:ldet}
The secrecy capacity $C_s$ of the linear deterministic wiretap channel \eqref{eq:ch-lsdpc} with side information is given by
\begin{equation}
C_s = \left\{
\begin{array}{l}
n_1, \hspace{55pt} \mbox{if $n_1-m_1 \neq n_2-m_2$,}\\
\hspace{80pt} \mbox{$n_1 \leq m_1$ or $n_2 \leq m_2$}\\
\max\left\{m_1,n_1-n_2+m_2\right\},\\
\hspace{70pt} \mbox{if $n_1-m_1 \neq n_2-m_2$,}\\
\hspace{80pt} \mbox{$n_1 > m_1$ and $n_2 > m_2$}\\
(n_1-n_2)^+, \hspace{18pt} \mbox{if $n_1-m_1 = n_2-m_2$}.
\end{array}
\right.\label{eq:ldet2}
\end{equation}
\end{theorem}

To prove Theorem~\ref{thm:ldet}, let us first prove the following proposition.

\begin{proposition}\label{prop:ldet}
The secrecy capacity $C_s$ of the linear deterministic wiretap channel \eqref{eq:ch-lsdpc} with side information is given by
\begin{equation}
C_s =  \min\left\{n_1,
rank\left(
\left[
\begin{array}{l}
A\\ B
\end{array}\right]\right)-rank(B)\right\}
\label{eq:ldet}
\end{equation}
where 
\begin{equation}
\begin{array}{rcl}
A & := & \left[
\begin{array}{ll}
D^{q-n_1} & D^{q-m_1}
\end{array}\right]\\
\mbox{and} \quad B & := & \left[
\begin{array}{ll}
D^{q-n_2} & D^{q-m_2}
\end{array}\right].
\end{array}
\label{eq:AB}
\end{equation}
\end{proposition}

\begin{IEEEproof}
To prove \eqref{eq:ldet}, we shall show that for the linear deterministic model \eqref{eq:ch-lsdpc}, both $H(Y_1|S)$ and $H(Y_1|Y_2)$ are simultaneously maximized when $X$ is an i.i.d. Bernoulli-$1/2$ random vector and independent of $S$.

First,
\begin{eqnarray}
H(Y_1|S) &=& H(D^{q-n_1}X|S)\nonumber\\
& \leq & H(D^{q-n_1}X)\nonumber\\
& \leq & rank(D^{q-n_1})\nonumber\\
& =& n_1\label{eq:t100}
\end{eqnarray}
where the equalities hold when $X$ is an i.i.d. Bernoulli-$1/2$ random vector and independent of $S$. 

To show that $H(Y_1|Y_2)$ is also maximized when $X$ is an i.i.d. Bernoulli-$1/2$ random vector and independent of $S$, we shall need the following technical lemma, which can be proved using a counting argument as provided in Appendix~\ref{app2}.

\begin{lemma}\label{lemma:ent}
For any matrices $A$ and $B$ in $\mathbb{F}_2$ (Galois field of size 2) that have the same number of columns,
\begin{equation}
\max H(AZ|BZ) = rank\left(\left[
\begin{array}{l}
A\\ B
\end{array}\right]
\right)-rank(B)
\end{equation} 
where the maximization is over all possible binary random vector $Z$. The maximum is achieved when $Z$ is an i.i.d. Bernoulli-$1/2$ random vector.
\end{lemma}

Now let 
$$Z := 
\left[
\begin{array}{c}
  X   \\
  S
\end{array}
\right].$$ 
By Lemma~\ref{lemma:ent}, 
\begin{eqnarray}
H(Y_1|Y_2) & = & H(AZ|BZ)\nonumber\\
& \leq & rank\left(\left[
\begin{array}{l}
A\\ B
\end{array}\right]
\right)-rank(B)
\label{eq:t200}
\end{eqnarray}
where the equality holds when $X$ is an i.i.d. Bernoulli-$1/2$ random vector and independent of $S$. 

Substituting \eqref{eq:t100} and \eqref{eq:t200} into \eqref{eq:bd} completes the proof of the proposition.
\end{IEEEproof}

Given Proposition~\ref{prop:ldet}, the explicit characterization \eqref{eq:ldet2} of the secrecy capacity $C_s$ can be obtained from \eqref{eq:ldet} by evaluating the rank of the matrices 
\[\left[
\begin{array}{l}
A\\ B
\end{array}\right]\]
and $B$. The details of the evaluation process are provided in Appendix~\ref{app3}.

\subsection{Degraded Gaussian Model}
Finally, let us consider the Gaussian wiretap channel where the received signals (at time index $i$) at the legitimate receiver and the eavesdropper are given by
\begin{equation}
\begin{array}{lll}
Y_1[i] &=& h_1X[i]+g_1S[i]+N_1[i]\\
Y_2[i] &=& h_2X[i]+g_2S[i]+N_2[i].
\end{array}
\label{eq:ch-sdpc}
\end{equation}
Here, $X[i]$ is the channel input which is subject to a \emph{unit} average power constraint, $N_k[i]$, $k=1,2$ and $S[i]$ are independent \emph{standard} Gaussian noise and interference and are i.i.d. across the time index $i$, and $h_1$, $h_2$, $g_1$ and $g_2$ are the (real) channel coefficients. The interference $S[i]$ is assumed to be non-causally known at the transmitter as side information. The channel coefficients $h_1$, $h_2$, $g_1$ and $g_2$ are fixed during communication and are assumed to be known at all terminals. 

A single-letter expression for an achievable secrecy rate was given in \eqref{eq:lb}, which involves an auxiliary variable $U$. However, it is \emph{not} clear what would be a reasonable choice of $U$, letting alone finding an optimal one that maximizes the achievable secrecy rate expression \eqref{eq:lb}. On the other hand, for the linear deterministic model \eqref{eq:ch-lsdpc}, it is clear from Theorem~\ref{thm:det} and Proposition~\ref{prop:ldet} that the following choice of auxiliary variable $U$ is optimal: 
\begin{equation}
U=Y_1=D^{q-n_1}X \oplus D^{q-m_1}S
\label{eq:aux3}
\end{equation}
where $X$ is an i.i.d. Bernoulli-$1/2$ random vector and independent of $S$. 

Based on the equivalence relationship \eqref{eq:eqv} between the Gaussian and the linear deterministic model and the success of Sec.~\ref{sec:dpc} for Costa's dirty-paper channel, the optimal choice \eqref{eq:aux3} of auxiliary variable $U$ for the linear deterministic model \eqref{eq:ch-lsdpc} suggests the following choice of auxiliary variable $U$ for the Gaussian model \eqref{eq:ch-sdpc}:
\begin{equation}
U=h_1X+g_1S
\label{eq:aux4}
\end{equation}
where $X$ is standard Gaussian and independent of $S$. For this choice of auxiliary-input variable pair $(U,X)$, 
\begin{eqnarray}
\hspace{-20pt} I(U;S) 
& = & \frac{1}{2}\log\left(1+\frac{g_1^2}{h_1^2}\right)\label{eq:GX}\\
\hspace{-20pt} I(U;Y_1) 
&=& \frac{1}{2}\log(1+h_1^2+g_1^2)\\
\hspace{-20pt} \mbox{and} \quad I(U;Y_2) 
&=& \frac{1}{2}\log\frac{(h_1^2+g_1^2)(1+h_2^2+g_2^2)}{h_1^2+g_1^2+(h_1g_2-h_2g_1)^2}
\label{eq:Ste}
\end{eqnarray}
giving
\begin{equation*}
I(U;Y_1)-I(U;S) = \frac{1}{2}\log\frac{(1+h_1^2+g_1^2)h_1^2}{h_1^2+g_1^2} 
\end{equation*}
and
\begin{eqnarray*}
\hspace{-20pt} && I(U;Y_1)-I(U;Y_2)\\
\hspace{-20pt} && \hspace{10pt}  = \; \frac{1}{2}\log\frac{(1+h_1^2+g_1^2)[h_1^2+g_1^2+(h_1g_2-h_2g_1)^2]}{(h_1^2+g_1^2)(1+h_2^2+g_2^2)}.
\end{eqnarray*}
By the single-letter achievable secrecy rate expression \eqref{eq:lb},  
\begin{equation}
\begin{array}{l}
R_s=\left(\min\left\{\frac{1}{2}\log\frac{(1+h_1^2+g_1^2)h_1^2}{h_1^2+g_1^2},\right.\right.\\
\hspace{40pt}\left.\left.\frac{1}{2}\log\frac{(1+h_1^2+g_1^2)[h_1^2+g_1^2+(h_1g_2-h_2g_1)^2]}{(h_1^2+g_1^2)(1+h_2^2+g_2^2)}\right\}\right)^+
\label{eq:Rs}
\end{array}
\end{equation}
is an achievable secrecy rate for the Gaussian wiretap channel \eqref{eq:ch-sdpc} with side information.

Following the works \cite{MVL-IT06} and \cite{CV-IT08}, below we focus on the special case where 
\begin{equation}
h_2=\beta h_1 \quad \mbox{and} \quad g_2 =\beta g_1
\label{eq:deg}
\end{equation}
for some $|\beta| \leq 1$. Note that the secrecy capacity of the channel \eqref{eq:ch-sdpc} does \emph{not} depend on the correlation between the additive Gaussian noise $N_1[i]$ and $N_2[i]$, so we may write  
$$N_2[i]=\beta N_1[i]+N[i]$$ 
where $N[i]$ is Gaussian with zero mean and variance $1-\beta^2$ and independent of $N_1[i]$. Thus, for the special case of \eqref{eq:deg}, the channel \eqref{eq:ch-sdpc} can be equivalently written as
\begin{equation}
\begin{array}{lll}
Y_1[i] &=& h_1X[i]+g_1S[i]+N_1[i]\\
Y_2[i] &=& \beta Y_1[i]+N[i]
\end{array}
\label{eq:ch-sdpc2}
\end{equation}
i.e., the received signal $Y_2[i]$ at the eavesdropper is \emph{degraded} with respect to the the received signal $Y_1[i]$ at the legitimate receiver. 

Following \cite{Cos-IT83}, an interesting interpretation of the degraded Gaussian model \eqref {eq:ch-sdpc2} is ``secret writing on dirty paper." In this scenario, a user intends to convey (to a legitimate receiver) a confidential message on a piece of paper with preexisting dirt on it. The legitimate receive has access to the \emph{original} paper with the message written on it and hence can decode the intended message. On the other hand, the eavesdropper can only access a \emph{noisy} copy of the original paper, from which essentially no information on the conveyed message can be inferred.

Next, we show that for the degraded Gaussian model \eqref{eq:ch-sdpc2}, the achievable secrecy rate \eqref{eq:Rs} is always within half a bit of the secrecy capacity. The result is summarized in the following theorem.

\begin{theorem}\label{thm:app}
For the degraded Gaussian wiretap channel \eqref{eq:ch-sdpc2} with side information, the secrecy capacity $C_s$ can be bounded as
\begin{equation}
\begin{array}{l}
\left(\min\left\{\frac{1}{2}\log\frac{(1+h_1^2+g_1^2)h_1^2}{h_1^2+g_1^2},\frac{1}{2}\log \frac{1+h_1^2+g_1^2}{1+\beta^2(h_1^2+g_1^2)}\right\}\right)^+ \leq \\
\hspace{40pt} C_s \leq \min\left\{\frac{1}{2}\log(1+h_1^2),\frac{1}{2}\log\frac{2(h_1^2+g_1^2)+1}{2\beta^2(h_1^2+g_1^2)+1}\right\}.
\end{array}
\label{eq:app}
\end{equation}
Moreover, the lower bound here is always within half a bit of the upper bound.
\end{theorem}

\begin{IEEEproof}
The lower bound in \eqref{eq:app} follows from \eqref{eq:Rs} and the degradedness assumption \eqref{eq:deg}. To prove the upper bound, note that for any input variable $X$ such that $\mathbb{E}[X^2] \leq 1$ we have
\begin{eqnarray}
I(X;Y_1|S) & = & h(Y_1|S)-h(Y_1|X,S)\nonumber\\
& = & h(h_1X+N_1|S)-h(N_1)\nonumber\\
& \leq & h(h_1X+N_1)-h(N_1)\nonumber\\
& \leq & \frac{1}{2}\log(1+h_1^2\mathrm{Var}(X))\nonumber\\
& \leq & \frac{1}{2}\log(1+h_1^2).\label{eq:ub1}
\end{eqnarray}
Furthermore, 
\begin{eqnarray}
\hspace{-20pt} && I(X,S;Y_1|Y_2)\nonumber\\
\hspace{-20pt} && \hspace{20pt} = \; h(Y_1|Y_2)-h(Y_1|X,S,Y_2)\nonumber\\
\hspace{-20pt} && \hspace{20pt} = \; h(Y_1|\beta Y_1+N)-h(N_1|\beta N_1+N)\nonumber\\
\hspace{-20pt} && \hspace{20pt} = \; h(Y_1|\beta Y_1+N)-\frac{1}{2}\log\left(2\pi{e}(1-\beta^2)\right).\label{eq:t1}
\end{eqnarray}
By an inequality of Thomas \cite[Lemma~1]{Tho-IT87} and the independence between $Y_1$ and $N$,
\begin{equation}
h(Y_1|\beta Y_1+N) \leq \frac{1}{2}\log\frac{2\pi{e}\mathrm{Var}(Y_1)(1-\beta^2)}{\beta^2\mathrm{Var}(Y_1)+(1-\beta^2)}.
\label{eq:t2}
\end{equation}
Note that the right-hand side of \eqref{eq:t2} is a monotone increasing function of $\mathrm{Var}(Y_1)$, which can be bounded from above as
\begin{eqnarray*}
\mathrm{Var}(Y_1) &=& \mathrm{Var}(h_1X+g_1S+N_1)\\
&=& \mathrm{Var}(h_1X+g_1S)+1\nonumber\\
& \leq & 2\left(\mathrm{Var}(h_1X)+\mathrm{Var}(g_1S)\right)+1\\
& \leq & 2h_1^2+2g_1^2+1.
\end{eqnarray*}
Hence,
\begin{eqnarray}
\hspace{-20pt} &&h(Y_1|\beta Y_1+N)\nonumber\\
\hspace{-20pt} && \hspace{20pt} \leq \; \frac{1}{2}\log\frac{2\pi{e}(2h_1^2+2g_1^2+1)(1-\beta^2)}{\beta^2(2h_1^2+2g_1^2+1)+(1-\beta^2)}\nonumber\\
\hspace{-20pt} && \hspace{20pt} = \; \frac{1}{2}\log\frac{2\pi{e}(2h_1^2+2g_1^2+1)(1-\beta^2)}{2\beta^2(h_1^2+g_1^2)+1}.\label{eq:t3}
\end{eqnarray}
Substituting \eqref{eq:t3} into \eqref{eq:t1}, we have
\begin{equation}
I(X,S;Y_1|Y_2) \leq\frac{1}{2}\log\frac{2(h_1^2+g_1^2)+1}{2\beta^2(h_1^2+g_1^2)+1}.
\label{eq:ub2}
\end{equation}
Further substituting \eqref{eq:ub1} and \eqref{eq:ub2} into \eqref{eq:ub} establishes the upper bound in \eqref{eq:app}.

To show that the lower bound is always within half a bit of the upper bound, let us define
\begin{eqnarray*}
a &:=& \frac{1}{2}\log(1+h_1^2)\\
b &:=& \frac{1}{2}\log\frac{2(h_1^2+g_1^2)+1}{2\beta^2(h_1^2+g_1^2)+1}\\
c &:=& \frac{1}{2}\log\frac{(1+h_1^2+g_1^2)h_1^2}{h_1^2+g_1^2}\\
\mbox{and} \quad d &:=& \frac{1}{2}\log \frac{1+h_1^2+g_1^2}{1+\beta^2(h_1^2+g_1^2)}.
\end{eqnarray*}
We shall consider the following two cases separately.

Case 1: $h_1^2<1$. In this case,
\[a=\frac{1}{2}\log(1+h_1^2) <\frac{1}{2}\]
and the gap between the upper and the lower bound can be bounded from above as
\begin{equation}
\min\{a,b\}-\left(\min\{c,d\}\right)^+ \leq \min\{a,b\} \leq a < \frac{1}{2}.
\label{eq:g0}
\end{equation} 

Case 2: $h_1^2\geq 1$. In this case,
\begin{eqnarray}
a-c &=& \frac{1}{2}\log(1+h_1^2)-\frac{1}{2}\log\frac{(1+h_1^2+g_1^2)h_1^2}{h_1^2+g_1^2}\nonumber\\
& \leq & \frac{1}{2}\log(1+h_1^2)-\frac{1}{2}\log(h_1^2)\nonumber\\
& = & \frac{1}{2}\log\left(1+\frac{1}{h_1^2}\right)\nonumber\\
& \leq & \frac{1}{2}.
\label{eq:g1}
\end{eqnarray}
Note that for any channel parameters $h_1$, $g_1$ and $\beta$,
\begin{eqnarray}
b-d &=& \frac{1}{2}\log\frac{2(h_1^2+g_1^2)+1}{2\beta^2(h_1^2+g_1^2)+1}
-\frac{1}{2}\log \frac{1+h_1^2+g_1^2}{1+\beta^2(h_1^2+g_1^2)}\nonumber\\
&=& \frac{1}{2}\log\left[\frac{2(h_1^2+g_1^2)+1}{1+h_1^2+g_1^2}\cdot\frac{1+\beta^2(h_1^2+g_1^2)}{2\beta^2(h_1^2+g_1^2)+1}\right]\nonumber\\
& \leq & \frac{1}{2}\log\frac{2(h_1^2+g_1^2)+1}{1+h_1^2+g_1^2}\nonumber\\
& \leq & \frac{1}{2}\log\frac{2(h_1^2+g_1^2)+2}{1+h_1^2+g_1^2}\nonumber\\
& = & \frac{1}{2}\label{eq:g2}
\end{eqnarray}
and for any real scalers $a$, $b$, $c$ and $d$,
\begin{eqnarray}
&& \min\{a,b\}-\left(\min\{c,d\}\right)^+\nonumber\\
&& \hspace{20pt} \leq \; \min\{a,b\}-\min\{c,d\}\nonumber\\
&& \hspace{20pt} = \; \max\left\{\min\{a,b\}-c,\min\{a,b\}-d\right\}\nonumber\\
&& \hspace{20pt} = \; \max\{a-c,b-d\}. \label{eq:g3}
\end{eqnarray}
Substituting \eqref{eq:g1} and \eqref{eq:g2} into \eqref{eq:g3}, we have 
\begin{equation}
\min\{a,b\}-\left(\min\{c,d\}\right)^+ \leq \max\left\{\frac{1}{2},\frac{1}{2}\right\}=\frac{1}{2}.
\label{eq:g4}
\end{equation}

Combining the above two cases proves that the lower bound in \eqref{eq:app} is always within half a bit of the upper bound. This completes the proof of the theorem.
\end{IEEEproof}

\begin{figure*}[t!]
\centering
\includegraphics[width=\linewidth,draft=false]{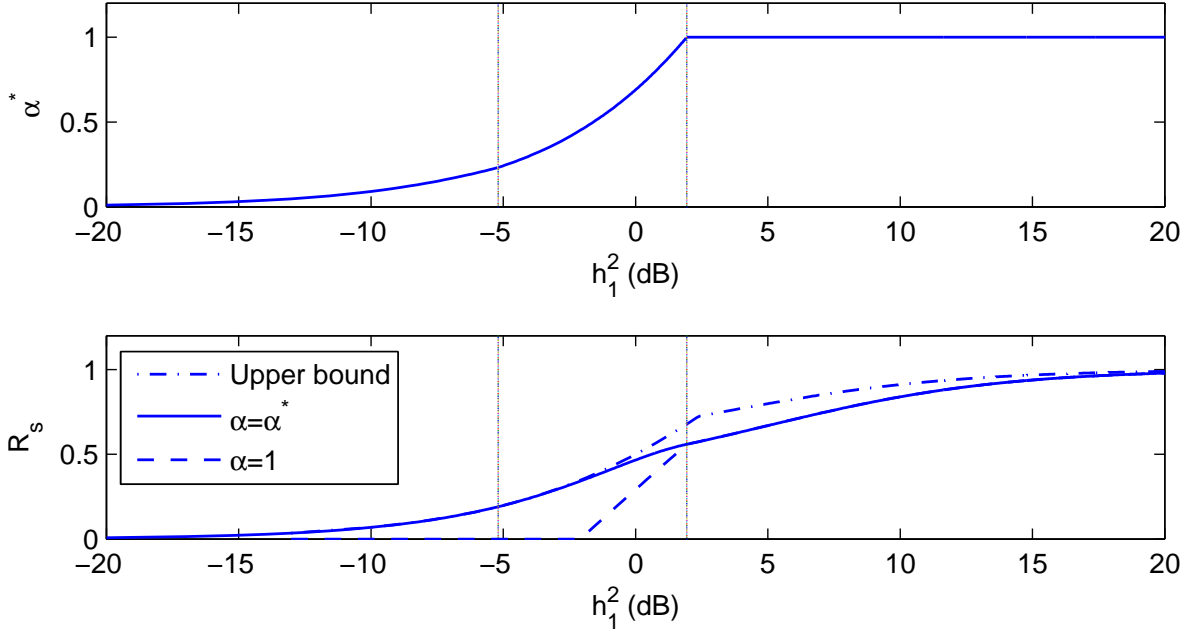}
\caption{A numerical comparison between the achievable secrecy rates for choosing $\alpha=\alpha^*$ and $\alpha=1$ in \eqref{eq:MVL}. Both $\alpha^*$ and the achievable secrecy rate $R_s$ are plotted as a function of $h_1^2$, while $g_1$ and $\beta$ are fixed to be $1$ and $0.5$, respectively.}
\label{fig2}
\end{figure*}

Finally, we note that the work \cite{MVL-IT06} considered, as a \emph{heuristic} choice, the auxiliary variable 
\begin{equation}
U=h_1X+\alpha g_1S \label{eq:MVL}
\end{equation}
where $X$ is standard Gaussian and independent of $S$, and $\alpha$ is chosen to maximize the achievable secrecy rate. A closed-form expression for the maximizing $\alpha$ can be written as
\[\alpha^*=\left\{
\begin{array}{ll}
\frac{h_1^2}{h_1^2+1}, & \mbox{if $0 \leq h_1^2 < h_{1L}^2$}\\
\frac{\beta^2h_{1}^{2}\left(|g_{1}|+\sqrt{h_{1}^{2}+g_{1}^{2}+1/\beta^{2}}\right)}{|g_{1}|(1+\beta^{2}h_{1}^{2})}, 
& \mbox{if $h_{1L}^2 \leq h_{1}^2 < h_{1H}^2$}\\
1, & \mbox{if $h_{1}^2 \geq h_{1H}^2$}\\
\end{array}
\right.\]
where
\begin{eqnarray*}
h_{1L}^{2} &=& \left(-\frac{g_{1}^{2}}{2}-1+\frac{|g_{1}|}{2}\sqrt{g_{1}^{2}+\frac{4}{\beta^2}-4}\right)^+\\
\mbox{and} \quad
h_{1H}^{2} &=& -\frac{g_{1}^{2}}{2}+\frac{|g_{1}|}{2}\sqrt{g_{1}^{2}+\frac{4}{\beta^2}}.
\end{eqnarray*}
Thus, for $h_1^2 \geq h_{1H}^2$, the heuristic choice \eqref{eq:MVL} with the maximizing $\alpha$ coincides with the choice $U=h_1 X+g_1S$ suggested by the linear deterministic model. 

A numerical comparison between the achievable secrecy rates for choosing $\alpha=\alpha^*$ and $\alpha=1$ in \eqref{eq:MVL} as well as the upper bound in \eqref{eq:app} is provided in Figure~\ref{fig2}. As we can see, when $h_1^2$ (which represents the received signal-to-noise ratio at the legitimate receiver) is small, the choice $\alpha=1$ (as suggested by the linear deterministic model) can be very suboptimal in maximizing the achievable secrecy rate. However, in this case, the secrecy capacity of the channel is also small, so the achievable secrecy rate given by the suboptimal choice $\alpha=1$ remains within half a bit of the secrecy capacity. For small $h_1^2$, substantial improvement to the achievable secrecy rate can be made by optimizing over $\alpha$. In fact, when $h_1^2 \leq h_{1L}^2$, the achievable secrecy rate given $\alpha=\frac{h_1^2}{1+h_1^2}$ coincides with the upper bound and hence gives the \emph{exact} secrecy capacity of the channel.  When $h_1^2$ is large, the maximizing $\alpha$ approaches $1$ (it is exactly equal to 1 when $h_1^2 \geq h_{1H}^2$), and both choices lead to achievable secrecy rates which are within half a bit of the secrecy capacity.

\section{Secret-Key Agreement via Dirty-Paper Coding}\label{sec:kdpc}
A different but closely related communication scenario is \emph{secret-key agreement} via dirty-paper coding, which was first considered in \cite{KDW-IFS11}. In this setting, the channel model is exactly the same as that for secret writing on dirty paper. The difference is in the goal of communication. For secret writing on dirty paper, the goal is to  convey to the legitimate receiver a secret message $W$, which is pre-chosen and hence is \emph{independent} of the known interference $\{S[i]\}$. For secret-key agreement, the goal is to establish, between the transmitter and the legitimate receiver, an agreement on a secret key $K$, which must be kept asymptotically perfectly secret from the eavesdropper, i.e., 
\[\frac{1}{n}I(K;Y_2^n) \rightarrow 0\]
in the limit as the block length $n \rightarrow \infty$.
The secret-key capacity $C_{K}$ is defined as the largest entropy rate $(1/n)\log H(K)$ that can be achieved by a coding scheme. Unlike the problem of secret writing on dirty paper, the secret key $K$ can be potentially \emph{correlated} with the known interference $\{S[i]\}$. Hence, the secret-key capacity $C_{K}$ is at least as large as the secrecy capacity $C_s$ for the same wiretap channel.

For a general discrete memoryless wiretap channel with side information, the secret-key capacity $C_K$ is unknown. The following lower and upper bounds were established in \cite{KDW-IFS11}:
\begin{equation}
\max_{p(u,x|s)} \left[I(U;Y_1)-I(U;Y_2)\right] \leq C_K \leq \max_{p(x|s)}I(X,S;Y_1|Y_2)
\label{eq:SK1}
\end{equation}
where $U$ is an auxiliary variable satisfying the Markov chain $U \rightarrow (X,S) \rightarrow (Y_1,Y_2)$ and such that
\begin{equation}
I(U;Y_1)-I(U;S) \geq 0.
\label{eq:SK2}
\end{equation} 
For semi-deterministic wiretap channels where the received signal $Y_1$ at the legitimate receiver is a deterministic bivariate function of the channel input $X$ and state $S$, the lower bound in \eqref{eq:SK1} with the choice of auxiliary variable $U=Y_1$ coincides with the upper bound, giving an \emph{exact} characterization of the secret-key capacity
\begin{equation}
C_K=\max_{p(x|s)}H(Y_1|Y_2).
\label{eq:SK3}
\end{equation}
Note here that the choice $U=Y_1$ always satisfies the constraint \eqref{eq:SK2}.

For the linear deterministic wiretap channel \eqref{eq:ch-lsdpc} with side information, by Lemma~\ref{lemma:ent} the conditional entropy $H(Y_1|Y_2)$ is maximized when the input variable $X$ is standard Gaussian and independent of $S$. By the equivalence relationship \eqref{eq:eqv} between the linear deterministic and the Gaussian model, this suggests the following choice of auxiliary variable $U$ for the degraded Gaussian model \eqref{eq:ch-sdpc2}:
\begin{equation}
U=h_1X+g_1S
\label{eq:SK4}
\end{equation}
where $X$ is standard Gaussian and independent of $S$, as long as \eqref{eq:SK2} is satisfied. Substituting \eqref{eq:SK4}, \eqref{eq:GX}--\eqref{eq:Ste}, and the degradedness assumption \eqref{eq:deg} into \eqref{eq:SK1} and \eqref{eq:SK2}, we have the following lower and upper bounds on the secret-key capacity $C_K$ of the degraded Gaussian model \eqref{eq:ch-sdpc2}:
\begin{equation}
\frac{1}{2}\log \frac{1+h_1^2+g_1^2}{1+\beta^2(h_1^2+g_1^2)} \leq C_K \leq \frac{1}{2}\log\frac{2(h_1^2+g_1^2)+1}{2\beta^2(h_1^2+g_1^2)+1}
\label{eq:SK5}
\end{equation}
for all channel coefficients $h_1$ and $g_1$ such that\footnote{The upper bound is valid for all channel parameters. Due to the constraint \eqref{eq:SK2}, when $h_1^2 < h_{1T}^2$ the linear deterministic model does not appear to provide any insight on how to choose the auxiliary variable $U$ for the degraded Gaussian model.}
\begin{equation}
h_1^2 \geq h_{1T}^2 := -\frac{g_1^2}{2}+\frac{|g_1|}{2}\sqrt{g_1^2+4}.
\label{eq:SK6}
\end{equation}
By \eqref{eq:g2}, the lower bound in \eqref{eq:SK5} is always within half a bit of the upper bound.


\begin{figure}[t!]
\centering
\includegraphics[width=\linewidth,draft=false]{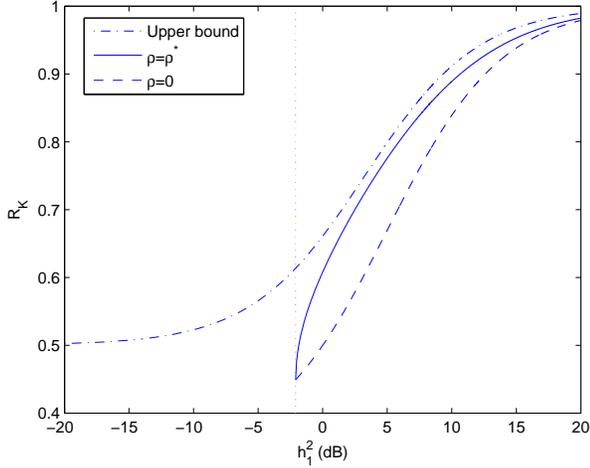}
\caption{A numerical comparison between the achievable secret-key rates for choosing $\rho=\rho^*$ and $\rho=0$ in \eqref{eq:SK4}. The achievable secret-key rate $R_K$ are plotted as a function of $h_1^2$, while $g_1$ and $\beta$ are fixed to be $1$ and $0.5$, respectively.}
\label{fig3}
\end{figure}

We mention here that \cite{KDW-IFS11} also considered, as a heuristic choice, the auxiliary variable $U$ of form \eqref{eq:SK4} where $X$ is standard Gaussian. However, instead of choosing $X$ to be independent of $S$ as suggested by the linear deterministic model, \cite{KDW-IFS11} considered $X$ which is correlated with $S$ and with correlation coefficient $\rho = \mathbb{E}[XS]=\rho^*$, where
\begin{equation}
\rho^*= \sqrt{1-\frac{h_1^2+g_1^2}{(1+h_1^2+g_1^2)h_1^2}}\cdot sgn(h_1g_1).
\label{eq:SK7}
\end{equation}
Here, $sgn(x)$ denotes the sign of real scalar $x$. It is straightforward to verify that condition \eqref{eq:SK6} guarantees the existence of $\rho^*$. A numerical comparison between the achievable secret-key rates for choosing the correlation coefficient $\rho=\rho^*$ and $\rho=0$ as well as the upper bound in \eqref{eq:SK5} is provided in Figure~\ref{fig3}. As we can see, even though the choice $\rho=0$ is suboptimal in maximizing the achievable secret-key rate, both choices lead to achievable secret-key rates that are within half a bit of the secret-key capacity for $h_1^2 \geq h_{1T}^2$.

\section{Concluding Remarks}\label{sec:con}
In this paper, we took a deterministic view and revisited the problem of wiretap channel with side information. A precise characterization of the secrecy capacity was obtained for a linear deterministic model, which naturally suggests a coding scheme which we showed to achieve the secrecy capacity of the degraded Gaussian model (dubbed as ``secret writing on dirty paper") to within half a bit. 

This paper falls in the line of using the linear deterministic model to provide approximate characterization of Gaussian network capacity, an approach which has become increasingly popular in information theory literature. However, our method is somewhat different from most of the practices along this line of research. In literature, a common practice has been to first gain ``insight" from the capacity-achieving scheme for the linear deterministic model and then translate the success to the Gaussian model at the \emph{scheme} level. To the best of our understanding, such translations are more art than science. For the problems that we considered in this paper, the translation of success from the linear deterministic model to the Gaussian model was done at the level of a \emph{single-letter description} of channel capacity and hence was much more systematic. Our ongoing work aims at understanding to what extent this method can be applied to more complex network communication scenarios.

%

\appendices

\section{Proof of Proposition~\ref{lemma:ub}}\label{app1}
By Fano's inequality, any achievable secrecy rate $R_s$ must satisfy
\begin{eqnarray*}
n(R_s-\epsilon_n) & \leq & I(W;Y_1^n)\\
& \leq &  I(W;Y_1^n,S^n)\\
& = &  I(W;Y_1^n|S^n)\\
& \leq &  I(X^n;Y_1^n|S^n)\\
& = & H(Y_1^n|S^n)-H(Y_1^n|X^n,S^n)\\
& = & H(Y_1^n|S^n)-\sum_{i=1}^nH(Y_1[i]|X[i],S[i])\\
& \leq & \sum_{i=1}^{n}H(Y_1[i]|S[i])-\sum_{i=1}^nH(Y_1[i]|X[i],S[i])\\
& = & n[H(Y_{1,Q}|S_Q,Q)-H(Y_{1,Q}|X_Q,S_Q,Q)]\\
& = & n[H(Y_{1,Q}|S_Q,Q)-H(Y_{1,Q}|X_Q,S_Q)]\\
& \leq & n[H(Y_{1,Q}|S_Q)-H(Y_{1,Q}|X_Q,S_Q)]\\
& = & n\cdot I(X_Q;Y_{1,Q}|S_Q)
\end{eqnarray*}
where $\epsilon_n \rightarrow 0$ in the limit as $n \rightarrow \infty$, and $Q$ is a standard time-sharing variable. 

Similarly, for any achievable secrecy rate $R_s$ we have
\begin{eqnarray*}
\hspace{-20pt} && n(R_s-\epsilon_n)\\
\hspace{-20pt} && \hspace{20pt} \leq \; I(W;Y_1^n)-I(W;Y_2^n)\\
\hspace{-20pt} && \hspace{20pt} \leq \;  I(W;Y_1^n,Y_2^n)-I(W;Y_2^n)\\
\hspace{-20pt} && \hspace{20pt} = \;  I(W;Y_1^n|Y_2^n)\\
\hspace{-20pt} && \hspace{20pt} \leq \;  I(X^n,S^n;Y_1^n|Y_2^n)\\
\hspace{-20pt} && \hspace{20pt} = \; H(Y_1^n|Y_2^n)-H(Y_1^n|X^n,S^n,Y_2^n)\\
\hspace{-20pt} && \hspace{20pt} = \; H(Y_1^n|Y_2^n)-\sum_{i=1}^nH(Y_1[i]|X[i],S[i],Y_2[i])\\
\hspace{-20pt} && \hspace{20pt} \leq \; \sum_{i=1}^{n}H(Y_1[i]|Y_2[i])-\sum_{i=1}^nH(Y_1[i]|X[i],S[i],Y_2[i])\\
\hspace{-20pt} && \hspace{20pt} = \; n[H(Y_{1,Q}|Y_{2,Q},Q)-H(Y_{1,Q}|X_Q,S_Q,Y_{2,Q},Q)]\\
\hspace{-20pt} && \hspace{20pt} = \; n[H(Y_{1,Q}|Y_{2,Q},Q)-H(Y_{1,Q}|X_Q,S_Q,Y_{2,Q})]\\
\hspace{-20pt} && \hspace{20pt} \leq \; n[H(Y_{1,Q}|Y_{2,Q})-H(Y_{1,Q}|X_Q,S_Q,Y_{2,Q})]\\
\hspace{-20pt} && \hspace{20pt} = \; n\cdot I(X_Q,S_Q;Y_{1,Q}|Y_{2,Q}).
\end{eqnarray*}

Note that the channel states are memoryless, so $S_Q$ has the same distribution as $S[i]$ for any $i=1,\ldots,n$. The channel is also memoryless, so the conditional distribution of $(Y_{1,Q},Y_{2,Q})$ given $(X_Q,S_Q)$ is given by the channel transition probability $p(y_1,y_2|x,s)$. Letting $X_Q=X$, $S_Q=S$, $Y_{1,Q}=Y_1$, $Y_{2,Q}=Y_2$, and $n \rightarrow \infty$ completes the proof of the proposition.

\section{Proof of Lemma~\ref{lemma:ent}}\label{app2}
Let $Z$ be an i.i.d. Bernoulli-$1/2$ vector. We have
\begin{eqnarray*}
H(AZ|BZ) & = & H\left(\left[\begin{array}{l}
A\\ B
\end{array}\right]Z\right)-H(BZ)\\
& = & rank\left(\left[\begin{array}{l}
A\\ B
\end{array}\right]\right)-rank(B).
\end{eqnarray*}
We thus conclude that 
\begin{equation}
\max H(AZ|BZ) \geq rank\left(\left[
\begin{array}{l}
A\\ B
\end{array}\right]
\right)-rank(B).
\label{eq:t501}
\end{equation} 

To prove the reverse inequality, let us consider the null space of $B$ and its coset partition based on the null space of 
$$\left[
\begin{array}{l}
A \\ B
\end{array}
\right].$$
Fix $BZ=b$. Then, any solution $Z$ can be written as the sum of a particular solution $Z_p$ and a vector $Z_h$ in the null space of $B$. Note that all vectors $Z_h$ in the same coset of the null space of $B$ relative to the null space of 
$$\left[
\begin{array}{l}
A \\ B
\end{array}
\right]$$ 
give the \emph{same} value for $AZ_h$. Thus, the number of \emph{different} values that $AZ$ can take for any given value of $b$ equals the number of cosets in the null space of $B$, which is given by 
\begin{equation*}
2^{nullitiy(B)-nullity\left(\left[
\begin{array}{l}
A\\ B
\end{array}\right]
\right)}
=
2^{rank\left(\left[
\begin{array}{l}
A\\ B
\end{array}\right]
\right)-rank(B)}.
\end{equation*}
We thus conclude that 
\begin{equation}
\max H(AZ|BZ) \leq rank\left(\left[
\begin{array}{l}
A\\ B
\end{array}\right]
\right)-rank(B).
\label{eq:t502}
\end{equation} 

Combining \eqref{eq:t501} and \eqref{eq:t502} completes the proof of the lemma.

\section{Proof of Theorem~\ref{thm:ldet}}\label{app3}
The matrix $B$ is a horizontal stack of two down-shift matrices with rank $n_2$ and $m_2$, respectively. Since both sub-matrices are in reduced row-echelon form, it suffices to count the number of nonzero rows of $B$ to find its rank:
\begin{eqnarray*}
rank(B) &=& q-\min\{q-n_{2},q-m_{2}\}\\
&=&\max\{n_{2},m_{2}\}.
\end{eqnarray*}

The matrix  
\[G:=\left[
\begin{array}{l}
A\\ B
\end{array}\right]=
\left[
\begin{array}{ll}
D^{q-n_1} & D^{q-m_1}\\ 
D^{q-n_2} & D^{q-m_2}
\end{array}\right]\]
is formed by vertically stacking two matrices $A$ and $B$. Thus, evaluating the rank of $G$ is equivalent to counting the number of zero rows along with the number of redundant nonzero rows between $A$ and $B$ (denoted by $d_{AB}$): 
\begin{eqnarray*}
rank(G) &=&  2q-\min\{q-n_{1},q-m_{1}\}-\\
&& \hspace{20pt} \min\{q-n_{2},q-m_{2}\}-d_{AB}\\
&=& \max\{n_{1},m_{1}\}+\max\{n_{2},m_{2}\}-d_{AB}.
\end{eqnarray*}

To calculate $d_{AB}$, let us consider the following five cases separately:

Case 1: Either $n_{1} \leq m_{1}$ and $n_{2} > m_{2}$, or $n_{2} \leq m_{2}$ and $n_{1} > m_{1}$. In this case, all nonzero rows of $G$ are independent so $d_{AB}=0$. By Proposition~\ref{prop:ldet},
\begin{equation*}
C_s = \min\{n_1,\max\{n_{1},m_{1}\}-0\}=n_1.
\end{equation*}

Case 2: $n_{1} \leq m_{1}$ and $n_{2} \leq m_{2}$, but $m_{1}-n_{1} \neq m_{2}-n_{2}$. In this case, the redundant nonzero rows of $G$ are given by the redundant rows between the top $m_1-n_1$ nonzero rows of $A$ and the top $m_2-n_2$ nonzero rows of $B$. Hence, $d_{AB}=\min\{m_{1}-n_{1},m_{2}-n_{2}\}$. By Proposition~\ref{prop:ldet},
\begin{eqnarray*}
C_s &=& \min\{n_{1},m_1-\min\{m_{1}-n_{1},m_{2}-n_{2}\}\}\\
&=&\min\{n_{1},\max\{n_{1},m_1-m_{2}+n_{2}\}\}\\
&=& n_1.
\end{eqnarray*}

Case 3: $n_{1} > m_{1}$ and $n_{2} > m_{2}$, but $m_{1}-n_{1} \neq m_{2}-n_{2}$. In this case, the redundant nonzero rows of $G$ are given by the redundant rows between the top $n_1-m_1$ nonzero rows of $A$ and the top $n_2-m_2$ nonzero rows of $B$. Hence, $d_{AB}=\min\{n_{1}-m_{1},n_{2}-m_{2}\}$. By Proposition~\ref{prop:ldet},
\begin{eqnarray*}
C_s &=& \min\{n_{1},n_1-\min\{n_{1}-m_{1},n_{2}-m_{2}\}\}\\
&=&n_{1}-\min\{n_{1}-m_{1},n_{2}-m_{2}\}\\
&=& \max\{m_1,n_1-n_2+m_2\}.
\end{eqnarray*}

Case 4:  $n_{1}-m_{1}= n_{2}-m_{2}$ and $n_{1}\geq m_{1}$. In this case, the redundant nonzero rows of $G$ correspond to the redundant nonzero rows of 
\[\left[
\begin{array}{l}
D^{q-n_1}\\ D^{q-n_2}
\end{array}\right]\]
so $d_{AB}=\min\{n_{1},n_{2}\}$. By Proposition~\ref{prop:ldet},
\begin{eqnarray*}
C_s &=& \min\{n_1,n_{1}-\min\{n_{1},n_{2}\}\}\\
&=& n_{1}-\min\{n_{1},n_{2}\}\\
&=& (n_1-n_2)^+.
\end{eqnarray*}

Case 5:  $n_{1}-m_{1}= n_{2}-m_{2}$ and $n_{1} < m_{1}$. In this case, the redundant nonzero rows of $G$ correspond to the redundant nonzero rows of 
\[\left[
\begin{array}{l}
D^{q-m_1}\\ D^{q-m_2}
\end{array}\right]\]
so $d_{AB}=\min\{m_{1},m_{2}\}$. By Proposition~\ref{prop:ldet},
\begin{eqnarray*}
C_s &=& \min\{n_1,m_{1}-\min\{m_{1},m_{2}\}\}\\
&=& \min\{n_1,(m_{1}-m_{2})^+\}\\
&=& \min\{n_1,(n_{1}-n_{2})^+\}\\
&=& (n_1-n_2)^+.
\end{eqnarray*}

Combining the results from the above five cases completes the proof of \eqref{eq:ldet2} and hence Theorem~\ref{thm:ldet}.

\section*{Acknowledgment}
The authors wish to thank the Associate Editor, Dr. Suhas Diggavi, for bringing the problem of secret-key agreement via dirty-paper coding and reference \cite{KDW-IFS11} to our attention after the initial submission of this paper.

\end{document}